\documentclass[a4paper, reqno]{amsart}
\usepackage{a4,wasysym}
\usepackage{amsmath,graphicx, amssymb,color,ulem,bbm}
\usepackage{float}
\newtheorem{theorem}{Theorem}

\newtheorem{proposition}[theorem]{Proposition}

\newcommand{\ud}{\mathrm{d}}

\newcommand{\eofproof}{\hfill$\Box$ }

\begin{document}

\title[Stability of heterogeneous predator-prey systems in space]{Revisiting the stability of spatially heterogeneous predator-prey systems under eutrophication}

\author[J\'{o}zsef Z. Farkas]{J. Z. Farkas}
\address{J\'{o}zsef Z. Farkas, Division of Computing Science and Mathematics, University of Stirling, Stirling, FK9 4LA, UK}
\email{jozsef.farkas@stir.ac.uk}

\author[A. Yu Morozov]{A. Yu Morozov}
\address{A. Yu Morozov, Department of Mathematics, University of Leicester, Leicester, LE1 7RH, UK}
\email{am379@leicester.ac.uk}

\author[E. G. Arashkevich]{E. G. Arashkevich}
\address{E. G. Arashkevich, Shirshov Institute of Oceanology, Moscow 117851, Russia}
\email{aelena@ocean.ru}

\author[A. Nikishina]{A. Nikishina}
\address{A. Nikishina, Shirshov Institute of Oceanology, Moscow 117851, Russia}
\email{anastasia.nikishina@gmail.com}

\subjclass{ }
\date{\today}

\begin{abstract}

We employ partial integro-differential equations to model trophic interaction in a spatially extended heterogeneous environment.  Compared to classical reaction-diffusion models, this framework allows us to more realistically describe the situation where movement of individuals occurs on a faster time scale than the demographic (population) time scale, and we cannot determine population growth based on local density. However, most of the results reported so far for such systems have only been verified numerically and for a particular choice of model functions, which obviously casts doubts  about these findings. In this paper, we analyse a class of  integro-differential predator-prey models with a highly mobile predator in a heterogeneous environment, and we reveal the main factors stabilizing such systems. In particular, we explore an ecologically relevant case of interactions in a highly eutrophic environment, where the prey carrying capacity can be formally set to 'infinity'. We investigate two main scenarios: (i) the spatial gradient of the growth rate is due to abiotic factors only, and (ii) the local growth rate depends on the global density distribution across the environment (e.g. due to non-local self-shading). For an arbitrary spatial gradient of the prey growth rate, we analytically investigate the possibility of the predator-prey equilibrium in such systems and we explore the conditions of stability of this equilibrium. In particular, we demonstrate that for a Holling type I (linear) functional response, the predator can stabilize the system at low prey density even for an 'unlimited' carrying capacity. We conclude that the interplay between spatial heterogeneity in the prey growth and fast displacement of the predator across the habitat works as an efficient stabilizing mechanism. These results highlight the generality of the  stabilization mechanisms we find in spatially structured predator-prey ecological systems in a heterogeneous environment.

\end{abstract}

\maketitle

\section{Introduction}

Understanding top-down control in food webs has been always a focus of theoretical ecology, and revealing potential mechanisms which stabilize predator-prey/resource-consumer interactions in ecosystems with a high degree of eutrophication (i.e. ecosystems that are high in nutrients) is a particularly challenging problem \cite{Rosenzweig_1971,Oaten1975, Armstrong_1994, Jansen_1995, Abrams_walters, Briggs2004, Morozov_etal2013}. Classical theory predicts that a generic predator-prey system will be highly unstable under eutrophic conditions and is expected to exhibit large-amplitude oscillations of species densities which will eventually result in population collapse and extinction \cite{Rosenzweig_1971, Gilpin1972}. However, these theoretical conclusions are often at odds with reality, since there exist a large number of examples of ecosystems in which the species densities remain low despite a high nutrient supply \cite{Chavez_etal1990, cullen_etal, Boyd2002, Jensen2005, Roy2007}. Various mechanisms have been proposed to explain such stability of predator-prey interactions in eutrophic environments. In particular, it has been suggested that this could be due to interplay between structuring within the prey population according to vulnerability or nutrition properties and active food selectivity of the predator (prey switching) \cite{Abrams_walters, Genkai-Kato_1999, Mougi_Nishimura2007,Morozov_etal2007, Morozov_etal2013, Farkas_Morozov_2014}. However, stabilization mechanisms of this kind cease working for unstructured populations, so a natural question is whether efficient top-down control is possible for a population of identical prey individuals.

Spatial heterogeneity of the environment and a non-homogeneous distribution of individuals across the habitat has been suggested to be an important factor promoting top-down control and preventing species extinction \cite{Jansen_1995, Nisbet_1997, Poggiale_Auger_2004, Briggs2004, Petrovskii2004}. In recent model studies of planktonic systems with a high nutrient supply, it was shown that interplay between the gradient of light in the water column and fast vertical displacement of herbivorous zooplankton can stabilize the system even for an ‘unlimited’ nutrient stock  \cite{Morozov_2011, Lewis2013}. The works cited implement a modelling approach which differs from the classical reaction-diffusion framework in several important ways. In particular, the new approach allows for different time scales for the population growth rate of prey and its predator and also takes into account the fact that the rate  at which predator disperses across the habitat can be substantially faster than its characteristic demographic scale. For example, zooplankton herbivores can quickly move vertically along the entire habitat (in the euphotic part of water column) a large number of times between reproduction events, thus we cannot assign an organism to a particular spatial location \cite{Cottier2006}, as in the case of reaction-diffusion models. Thus we should describe the predator population as a whole in terms of the total population size, while still taking into account non-homogeneity of the distribution of predator individuals across the habitat when considering grazing of the prey at each particular location. Ecologically similar scenarios can be found in a large number of other ecosystems when modelling, for instance, interactions between terrestrial herbivorous mammals and grass, zooplankton and fish, insects and birds, insects and their parasitoids, acarine predator-prey systems and even between fish population and fishing boats \cite{Milinski1979, Godin1984, Lodge1988, Kacelnik1992, Begon2005, Nachman2006, Auger2010}.

Modelling predator-prey interaction in a spatially heterogeneous habitat, where an exact location of individuals on a demographic timescale  cannot be properly assigned, requires the implementation of  \emph{partial integro-differential equations}. However, the complexity of this framework makes it rather difficult to treat the model analytically: all previous findings have been obtained by direct numerical simulations of the equations for particular parameterisations of the model ingredients \cite{Morozov_2011, Lewis2013}. This, obviously, cannot be considered as a rigorous proof of stabilization of the system. Moreover, the results can strongly depend on the choice of the per capita population growth rate, which in the previous studies was considered to be an exponential function modelling light attenuation with depth \cite{Morozov_2011, Lewis2013}. Finally, we should say that the integro-differential equations in \cite{Morozov_2011, Lewis2013} differ significantly from those used in the modelling of physiologically structured populations, see e.g. \cite{Cushing1998, Farkas_Hagen2007}. For instance, the model we develop here does not contain a transport term as standard physiologically structured models do.

In this paper, we revisit the previous results on stabilization for a class of integro-differential predator-prey models, and explore the main factors assuring efficient top-down control. We analytically obtain criteria for the existence of a positive stationary state assuring the coexistence of both species for an arbitrary gradient of resource distribution across the environment. Then we investigate the stability of the coexistence equilibrium and derive a general characteristic equation, which governs the stability of the system. We analytically show that in the case that there is no saturation in the functional response (i.e. Holling type I response), the predator is able to stabilize the system at low prey density even for an 'unlimited' carrying capacity. We study how the size of the habitat and the spatial gradient of the prey growth affects stabilization in the system. We explore two ecologically relevant scenarios: (i) the case when the spatial gradient of the growth rate of the prey is due to abiotic factors only, and (ii) the case where the growth rate of the prey is affected by the surrounding density of the population itself (e.g. effects of self-shading). We find that, interestingly, in case (i) the trophic control of the prey is not possible when the population size of the predator varies slowly, whereas in case (ii) the predator can efficiently suppress the population growth even without changing its own population size.

\section{Modelling framework and biological rationale}

The general predator-prey model with a highly mobile predator is described by the following system of partial integro-differential equations \cite{Morozov_2011}.
\begin{align}
\frac{\partial p}{\partial t}(h,t)&=D \frac{\partial^2 p}{\partial h^2}(h,t) + R(h,p(h,t))p(h,t)-f(p(h,t)) z(h,t),\quad p(h,0)=p_0(h), \label{prey} \\
\frac{\ud Z}{\ud t}(t)&=\frac{k}{H}\int_0^H f(p(h,t)) z(h,t)\,\ud h-mZ(t),\quad Z(0)=Z_0, \label{predator} \\
Z(t)&=\frac{1}{H}\int_0^H z(h,t)\,\ud h,\quad P(t)=\frac{1}{H}\int_0^H p(h,t)\,\ud h.\label{total_predator}
\end{align}
In the model above $p(h,t)$ and $z(h,t)$ denote the densities of prey and predator at time $t$ and at spatial location $h$, respectively; $H$ is the overall size of the habitat; hence $P(t)$ and $Z(t)$ are the average densities of the species at time $t$; while $p_0$ and $Z_0$ denote the initial conditions. The first term in the right hand side of \eqref{prey} describes the dispersal of prey across the habitat due to diffusion; $R$ is the function (or functional, in general) describing the local per capita growth rate of the prey; $f$ is the local functional response of the predator. The parameters $k$ and $m$ denote the food efficiency of the predator and its mortality rate, respectively. In the case where we model plankton interactions, $h$ is the vertical coordinate from the water surface and $H$ is the overall depth of the euphotic zone.

 Here we impose zero flux boundary conditions $\frac{\partial p}{\partial h}=0$ at  $h=0$ and $h=H$.

The integro-differential framework \eqref{prey}-\eqref{total_predator} is quite different from the conventional reaction-diffusion modelling setting. Here we consider that the predator movement occurs on a much faster timescale than both the prey movement and the population dynamics; and assume that the predator can quickly move to any part of the habitat. Thus, we cannot assign a particular spatial location to an individual within the predator population, since organisms constantly change their locations. The framework still allows us to consider the instantaneous spatial distribution of the predator population $z(h,t)$. Note that equations \eqref{prey}-\eqref{total_predator} do not uniquely specify $z(h,t)$ based on the knowledge of $Z(t)$, so we need to impose some further assumptions. A number of empirical observations suggest that the relative distribution of predator between patches often follows the relative abundance of food, which is described by
\begin{align}
z(h,t)=\frac{p(h,t)}{P(t)}Z(t). \label{z_distribution}
\end{align}
The above proportion-based parametrization may not be realistic since it does not take into account possible interference between predators \cite{Murdoch1992}. However, in a large number of study cases, see e.g. \cite {Milinski1979, Godin1984, Kacelnik1992, Larsson1997, Nachman2006},  we can still reasonably well approximate the instantaneous spatial distribution of predator across the food patches by \eqref{z_distribution}. For instance, the distribution of herbivorous zooplankton in the water column often follows that of phytoplankton, see e.g. \cite{Lampert2005, Morozov_2011}.

Fig.1a shows a typical example of the relative distribution of the biomass of phytoplankton (prey) and herbivorous zooplankton (predator) in different layers of the vertical water column in the euphotic zone of the sea. Here we split the water column into a number of layers and plot the values of the spatially averaged biomass of $p$ and $z$ in each layer divided by the overall average biomass across the entire column. The data were recently collected at 12 stations in the Black Sea for various seasons (2007-2013); further details on obtaining the samples are provided in electronic Appendix 1. The figure presents both night and day observations. For the night observation (open circles) one can see that the data can be roughly described by the straight line with slope approximately equal to 1, thus the basic assumption \eqref{z_distribution} holds (standard linear regression gives $R^2=0.85$ and the slope of the fitting line as 0.96).

The major difference between the night and the day patterns is that the vertical distribution of zooplankton substantially varies due to the phenomenon of regular daily vertical migration: during the day time zooplankton is mainly located in deep layers to avoid predation by  higher visual predators (e.g. fish) and ascend to upper layers at night \cite{Ohman1990}. Note that daytime feeding activity of herbivorous zooplankton is usually low and most of the food is consumed during the night time \cite{Bollens1989}.

Fig.1b provides examples of vertical distributions of species and the daily zooplankton migration pattern. One can see that the vertical distribution of zooplankton is highly variable throughout the 24 hour period. Moreover, zooplankton usually exhibit fast short term unsynchronized migrations \cite{Cottier2006}, by permanently descending and ascending over smaller distances (10-20m). Thus, we cannot assign a particular location to an individual zooplankton and we can use the integro-differential framework with an instantaneous distribution of the predator. We should note that model  \eqref{prey}-\eqref{total_predator} describes the organisms averaged over the day, and therefore excluding daily migration patterns. Finally, Fig.1b also provides the profile of chlorophyll $a$ through the column, which is a proxy of the distribution of phytoplankton. Its complicated structure is due to high heterogeneity in the distribution of the vital resources: light and nutrients (phosphorous and nitrogen) through the column, which potentially translates into a complicated relation $r(h)$ (cf. \cite{Ryabov}).

 We consider that the local functional response is given by the Holling type II function \cite{Holling1959, Gentleman_etla2003}
\begin{align}
f(p)=\alpha \frac{p}{1+p \beta},
\end{align}
where the parameters $\alpha$ and $\beta$ denote the attack rate and the inverse saturation food density, respectively. In the limiting case when $\beta=0$ the functional response becomes linear or Holling type I response. Here we intentionally implement a concave downwards (i.e. non-sigmoidal) functional response, which is a destabilizing factor in predator-prey models in general, see e.g. \cite{Rosenzweig_1971, Gilpin1972}.

For the sake of simplicity we neglect the effect of diffusion of the prey by setting $D=0$. In the Discussion section, we shall briefly address the effect that prey diffusion has on the results.

Finally, we need to specify the growth function $R$. Since we model eutrophic conditions, we consider that this function does not depend on the variation of nutrient concentration. We shall explore two main scenarios: (i) the per capita growth rate is a function of the abiotic environment only, i.e. it is a local function $R=r(h)$ and (ii) the per capita growth rate depends on the population density distribution in some parts of the environment $R=R(h, p)$, i.e. it is a non-local term. In the latter case, we have a functional, i.e. a function which depends on another function characterizing the overall spatial distribution of $p$. Note that in both cases, we neglect natural mortality of the prey compared to the predation rate.

 In case (i) the equations read (Model 1)
 \begin{align}
\frac{\partial p}{\partial t}(h,t)&=r(h)p(h,t)-\frac{1}{P(t)}\alpha\frac{p^2(h,t)}{1+\beta p(h,t)}Z(t),\quad p(h,0)=p_0(h), \label{prey1} \\
\frac{\ud Z}{\ud t}(t)&=k\frac{Z(t)}{HP(t)}\alpha\int_0^H\frac{p^2(h,t)}{1+\beta p(h,t)}\,\ud h-mZ(t),\quad Z(0)=Z_0. \label{predator1}
\end{align}
For the moment we do not specify the function $r$. We only assume that it is a smooth and non-negative function.

In case (ii), we shall investigate the particular scenario of self-shading in the population. In this complex scenario, the supply of a resource such as light decreases in space via a function of the local population density. In planktonic communities this may be shading of light for phytoplankton residing in deeper layers by the organisms dwelling in upper layers. For a non-planktonic system, this may be for instance a gradual depletion of a nutrient as it flows through part of the environment occupied by the prey and is gradually consumed by the organisms. In this case, the local rate of decrease of resource density can be assumed to be a function of the density of consumers. It is natural to assume that the
available resource quantity is a decreasing function of the total number of consumers $\int_0^h p(x,t)\,\ud x$  that are above/upstream an individual at depth/point $h$. In particular, we assume that the quantity of the available resource is an exponentially decreasing function of the number of consumers that contribute to the shading.

The equations read (Model 2)
\begin{align}
\frac{\partial p}{\partial t}(h,t)&=r_0 \exp\left(-\nu\int_0^h p(x,t)\,\ud x\right)p(h,t)-\frac{1}{P(t)}\alpha\frac{p^2(h,t)}{1+\beta p(h,t)}Z(t),\quad p(h,0)=p_0(h), \label{prey2} \\
\frac{\ud Z}{\ud t}(t)&=k\frac{Z(t)}{HP(t)}\alpha\int_0^H\frac{p^2(h,t)}{1+\beta p(h,t)}\,\ud h-mZ(t),\quad Z(0)=Z_0, \label{predator2}
\end{align}
where $\nu$ is the coefficient of self-shading. In both of the models above the average densities $P(t)$ and $Z(t)$ are given by equation \eqref{total_predator}.

In this paper, we are mostly interested in the existence and stability of the positive stationary state of \eqref{prey1}-\eqref{predator1} and \eqref{prey2}-\eqref{predator2}. It is worth mentioning that a spatially homogeneous predator-prey model \eqref{prey}-\eqref{total_predator} with perfect mixing (i.e. for $r(h)=const$ and $z(h)=const$) will be globally unstable \cite{Bazykin}, thus the main question is whether it is possible  to stabilize this system  by introducing spatial heterogeneity (i.e. struc\-tu\-ring the population) and allowing a non-homogeneous distribution of predators. It is also worth noting that the term describing prey growth in equation \eqref{prey2} has infinite dimensional nonlinearity, and that structured population models with such nonlinearity are notoriously difficult to analyse, see e.g. \cite{CF2012}, \cite{Farkas2009}.

\section{Results}

\subsection{Model 1: The local growth rate of the prey is a function of abiotic factors only}

\subsubsection{Existence of a non-trivial (coexistence) steady state}\

We look for a strictly positive steady state $(p_*,Z_*)$ of model \eqref{prey1}-\eqref{predator1}.
We assume that $r$ is a strictly positive continuous function with
$$\displaystyle\max_{h\in [0,H]} \{r(h)\} = r_M.$$ We characterise the existence of the positive steady state in the following
proposition.
\begin{proposition}
Assume that
\begin{equation}
\alpha k>m\beta \label{ssexistence}
\end{equation}
holds. Then model \eqref{prey1}-\eqref{predator1} has a unique positive steady state.
\end{proposition}
\noindent {\bf Proof.}
The steady state equations read
\begin{align}
r(h) & = \alpha\frac{Z_*}{P_*}\frac{p_*(h)}{1+\beta p_*(h)},\quad h\in [0,H], \label{steady1} \\
Z_*& = \frac{\alpha k}{m H}\int_0^H\frac{p_*^2(h)}{1+\beta p_*(h)}\,\ud h, \label{steady2} \\
P_* & = \frac{1}{H}\int_0^H p_*(h)\,\ud h. \label{steady3}
\end{align}
For positive values $(P_*,Z_*)$ which satisfy
\begin{equation}
\alpha Z_*>\beta r_MP_*,\label{ssregion}
\end{equation}
we obtain from equation \eqref{steady1}
\begin{equation}
p_*(h)=\frac{r(h)}{\alpha\frac{Z_*}{P_*}-r(h)\beta},\quad h\in [0,H].\label{steady4}
\end{equation}
We then substitute the steady state $p_*$ given by formula \eqref{steady4} into equations \eqref{steady2}-\eqref{steady3}.
After simplification we obtain:
\begin{align}
P_*=& \frac{1}{H}\int_0^H\frac{r(h)}{\alpha\frac{Z_*}{P_*}-r(h)\beta}\,\ud h,\label{steady5} \\
Z_*=& \frac{k}{mH}\int_0^H \frac{r^2(h)}{\alpha\frac{Z_*}{P_*}-r(h)\beta}\,\ud h.\label{steady6}
\end{align}
From equation \eqref{steady6} we obtain:
\begin{align}
Z_*=& \frac{k}{mH}\int_0^H\left[-\frac{1}{\beta}\frac{\alpha\frac{Z_*}{P_*}-r(h)\beta}{\alpha\frac{Z_*}{P_*}-r(h)\beta}r(h)+\frac{1}{\beta}\frac{\alpha\frac{Z_*}{P_*}}{\alpha\frac{Z_*}{P_*}-r(h)\beta)}r(h)\right]\,\ud h \nonumber \\
 =& -\frac{k}{\beta mH}\int_0^H r(h)\,\ud h+\frac{k}{m}\frac{\alpha Z_*}{\beta}, \label{steady7}
\end{align}
from which we obtain
\begin{equation}
Z_*=\frac{k}{\alpha k-m\beta}\frac{1}{H}\int_0^H r(h)\,\ud h. \label{steady8}
\end{equation}

We substitute this expression into equation \eqref{steady5} to obtain
\begin{equation}
1=\frac{1}{H}\int_0^H\frac{r(h)}{\frac{1}{H}\displaystyle\int_0^Hr(h)\,\ud h\frac{\alpha k}{\alpha k-m\beta}-r(h)\beta P_*}\,\ud h=:F(P_*).\label{steady9}
\end{equation}
Also note that \eqref{ssregion} together with \eqref{steady8} require the existence of a (unique) solution of \eqref{steady9} in
the interval

\begin{equation}
\left(0,\frac{\alpha k}{\alpha k-m\beta}\frac{1}{\beta r_M}\frac{1}{H}\int_0^H r(h)\,\ud h\right).\label{steady10}
\end{equation}

Note that $F(0)=\frac{\alpha k-m\beta}{\alpha k}<1$, and that

\begin{equation}
F(P_*)\to +\infty,\quad \text{as}\quad P_*\to\left(\frac{\alpha k}{\alpha k-m\beta}\frac{1}{\beta r_M}\frac{1}{H}\int_0^H r(h)\,\ud h\right)^-.
\end{equation}
Hence there is a solution of equation \eqref{steady9} in the required interval \eqref{steady10}. It is also shown
that on the interval \eqref{steady10} the function $F$ is strictly monotonically increasing, hence the solution is unique, and the proof is complete.
\eofproof

Note that, we may relax the assumption of strict positivity of $r$, and allow it to vanish at some values of $h$. Then
equation \eqref{steady4} shows that the steady state $p_*$ will also vanish at those values of $h$, but Proposition 3.1 still  holds.

\subsubsection{Stability of the coexistence steady state}\

The stability of the coexistence equilibrium is determined by the roots of a characteristic equation, which we derive in the current subsection.
The linearisation around the coexistence steady state $(p_*,Z_*)$ of model \eqref{prey1}-\eqref{predator1} is computed as
\begin{align}
\frac{\partial \Delta p}{\partial t}(h,t) =&r(h)\Delta p (h,t) \nonumber \\
 -&\alpha\left(\frac{p^2_*(h)}{P_*(1+\beta p_*(h))}\Delta Z(t)-\frac{p^2_*(h)Z_*}{P_*^2(1+\beta p_*(h))}\Delta P(t)+\frac{Z_*p_*(h)(2+\beta p_*(h))}{P_*(1+\beta p_*(h))^2}\Delta p (h,t)\right), \label{linear0} \\
\frac{\ud \Delta Z}{\ud t}(t) =& -m \Delta Z(t)+\frac{\alpha k}{HP_*}\Delta Z(t)\int_0^H\frac{p^2_*(h)}{1+\beta p_*(h)}\,\ud h-\frac{\alpha k Z_*}{HP_*^2}\Delta P(t)\int_0^H\frac{p^2_*(h)}{1+\beta p_*(h)}\,\ud h \nonumber \\
& + \frac{\alpha kZ_*}{HP_*}\int_0^H \frac{p_*(h)(2+\beta p_*(h))}{(1+\beta p_*(h))^2}\Delta p(h,t)\,\ud h,\quad \Delta P(t)=\frac{1}{H}\int_0^H \Delta p(h,t)\,\ud h. \label{linear00}
\end{align}

 We look for the solution in the form of $\Delta p (h,t)=w(h)\exp(\lambda t)$ and $\Delta Z (t)=U\exp(\lambda t)$. This leads to the eigenvalue problem ($W$ is the spatial average of $w$)
\begin{align}
& \left(\lambda-r(h)+\alpha\frac{Z_*}{P_*}\frac{p_*(h)(2+\beta p_*(h))}{(1+\beta p_*(h))^2}\right)w(h)= \nonumber \\
& \quad\quad\quad-\alpha\frac{p_*^2(h)}{P_*(1+\beta p_*(h))}U+\alpha\frac{Z_*}{P_*^2}\frac{p_*^2(h)}{1+\beta p_*(h)}W, \label{evproblem1} \\
& \lambda U=-mU \nonumber \\
& \quad\quad+\frac{\alpha k}{H}\int_0^H\left(\frac{p^2_*(h)}{P_*(1+\beta p_*(h))}U-\frac{p^2_*(h)Z_*}{P_*^2(1+\beta p_*(h))}W+\frac{Z_*p_*(h)(2+\beta p_*(h))}{P_*(1+\beta p_*(h))^2}w(h)\right)\,\ud h. \label{evproblem2}
\end{align}

Using \eqref{steady4} we recast equation \eqref{evproblem1} as follows
\begin{equation}
\left(\lambda+r(h)\left(1-\frac{P_*}{Z_*}\frac{\beta}{\alpha}r(h)\right)\right)w(h)=\frac{r^2(h)}{\alpha Z_*-P_*\beta r(h)}W-\frac{P_*r^2(h)}{Z_*\left(\alpha Z_*-P_*\beta r(h)\right)}U. \label{evproblem3}
\end{equation}
Note that \eqref{ssregion} implies that $\alpha Z_*-P_*r(h)>0$, for $h\in [0,H]$.
Next we multiply equation \eqref{evproblem1} by $\frac{k}{H}$, then integrate it from $0$ to $H$ and add it to equation \eqref{evproblem2} to obtain:
\begin{equation}
\lambda kW+\lambda U=k\overline{W}-mU,\label{evproblem4}
\end{equation}
where
\begin{equation}
\overline{W}=\frac{1}{H}\int_0^H r(h)w(h)\,\ud h.
\end{equation}
We  re-arrange equation \eqref{evproblem3} as
\begin{equation}
w(h)=\frac{r^2(h)}{k_1(\lambda,h)\left(\alpha Z_*-P_*\beta r(h)\right)}W-\frac{r^2(h)}{k_1(\lambda,h)\left(\alpha Z_*-P_*\beta r(h)\right)}\frac{P_*}{Z_*}U,\quad h\in[0,H], \label{evproblem6}
\end{equation}
where
\begin{equation*}
k_1(\lambda,h)=\lambda+r(h)\left(1-\frac{P_*}{Z_*}\frac{\beta}{\alpha}r(h)\right),\quad h\in [0,H].
\end{equation*}
Note that $k_1(\lambda,\cdot)\ne 0$ for $\lambda\in\mathbb{C}$, Re$(\lambda)\ge 0$.

Next we integrate equation \eqref{evproblem6} from $0$ to $H$ and divide by $H$ to obtain
\begin{equation}\label{evproblem7}
W=k_2(\lambda)W-k_2(\lambda)\frac{P_*}{Z_*}U,
\end{equation}
where
\begin{equation*}
k_2(\lambda)=\frac{1}{H}\int_0^H\frac{r^2(h)}{k_1(\lambda,h)\left(\alpha Z_*-P_*\beta r(h)\right)}\,\ud h.
\end{equation*}
We also multiply equation \eqref{evproblem6} by $\frac{r(h)}{H}$ and integrate from $0$ to $H$ to obtain
\begin{equation}\label{evproblem8}
\overline{W}=k_3(\lambda)W-k_3(\lambda)\frac{P_*}{Z_*}U,
\end{equation}
where
\begin{equation*}
k_3(\lambda)=\frac{1}{H}\int_0^H\frac{r^3(h)}{k_1(\lambda,h)\left(\alpha Z_*-P_*\beta r(h)\right)}\,\ud h.
\end{equation*}
Next we substitute into equation \eqref{evproblem4} the expression for $\overline{W}$ given by  \eqref{evproblem8}. This, together with equation \eqref{evproblem7}, yields a homogeneous system of two equations for the variables $(W,U)$ as follows.
\begin{align}
0= & k\left(k_3(\lambda)-\lambda \right)W+\left(-k\frac{P_*}{Z_*}k_3(\lambda)-\lambda-m\right)U, \nonumber \\
0= & (1-k_2(\lambda))W+k_2(\lambda)\frac{P_*}{Z_*}U.\label{evproblem9}
\end{align}
For any value of $\lambda\in\mathbb{C}$ with Re$(\lambda)\ge 0$ a non-trivial solution of system \eqref{evproblem9} yields a non-trivial solution $w$ via equation \eqref{evproblem6}. On the other hand, any eigenvalue $\lambda$ with Re$(\lambda)\ge 0$ necessarily satisfies equations \eqref{evproblem9} for same non-trivial $(W,U)$.

We summarize our result in the following theorem:
\begin{theorem}
The  non-trivial steady state $(p_*,Z_*)$ of model \eqref{prey1}-\eqref{predator1} is locally asymptotically stable if
every solution $\lambda$ of the characteristic equation below has negative real part.
\begin{equation}\label{chareq}
K(\lambda):=k\frac{P_*}{Z_*}k_3(\lambda)-k_2(\lambda)\left(\lambda\left(1+k\frac{P_*}{Z_*}\right)+m\right)+\lambda+m=0,
\end{equation}
where $k_2$ and $k_3$ are defined earlier. On the other hand, the steady state is unstable if there is at least one solution of \eqref{chareq} with positive real part.
\end{theorem}

In general, the derived characteristic equation \eqref{chareq} is complicated and determining its roots might require extensive numerical simulations which might take as much time as directly solving the system of differential equations. However, this equation can be rather useful for construction of a Hopf bifurcation curve: one can set $\lambda= i b$ ($b$ is a positive real number) and solve the resulting equation. In the Discussion section we shall provide an example of construction of such a curve for a particular parametrization of the function $r$.

Using the characteristic equation above we are able to  address the stability of the system in the case when the demographic timescale of the predator is much slower than that of the prey. In this case, we can assume the population size of the predator to be constant. The result is given by the following remark.

\begin{remark}
The positive steady state $(p_*,Z_*)$ is always unstable under perturbations in the first component (in the prey population) only.
We consider perturbations in the $p$-subspace only. Perturbations in the first component only lead to the characteristic equation
\begin{align}
1= & k_2(\lambda)=\frac{1}{H}\int_0^H\frac{r^2(h)}{k_1(\lambda,h)\left(\alpha Z_*-P_*\beta r(h)\right)}\,\ud h \nonumber \\
= &\frac{1}{H}\int_0^H\frac{r^2(h)}{\left(\lambda+r(h)\left(1-\frac{P_*}{Z_*}\frac{\beta}{\alpha}r(h)\right)\right)\left(\alpha Z_*-\beta P_*r(h)\right)}\,\ud h.   \label{evproblem11}
\end{align}
It is shown that $k_2(0)>1$, and clearly $k_2$ is a monotone decreasing function of $\lambda$ for $\lambda>0$.
Also we have that $\displaystyle\lim_{\lambda\to +\infty}k_2(\lambda)=0$, hence there exists a unique positive eigenvalue, and the steady state is unstable.
\end{remark}

\begin{remark} Another important conclusion we may draw from \eqref{chareq} is that for a homogeneous environment (i.e. $r(h) \equiv  const$) the system is always unstable provided $\beta>0$, since all eigenvalues will have positive real parts (we do not show here the corresponding characteristic equation for the sake of simplicity).
\end{remark}

\subsubsection{Stability  in the special case $\beta=0$}\

Next we investigate an important special case of \eqref{chareq} where we have a Holling type I functional response, i.e. there is no saturation rate in predation, $\beta=0$. The characteristic equation reduces to
\begin{align}
K(\lambda)= & m+\lambda+\frac{P_*}{Z_*}\frac{k}{H}\int_0^H\frac{r^3(h)}{(\lambda+r(h))\alpha Z_*}\,\ud h \nonumber \\
 & -\left(\lambda\left(1+k\frac{P_*}{Z_*}\right)+m\right)\frac{1}{H}\int_0^H\frac{r^2(h)}{(\lambda+r(h))\alpha Z_*}\,\ud h=0,\label{chareq2}
\end{align}
which can be simplified by recalling the expressions for the stationary densities of species \eqref{steady5}, \eqref{steady6} and \eqref{steady8}. We obtain
\begin{align}
K(\lambda)= & m-\frac{2m\lambda}{\int_0^H r^2(h) dh} \int_0^H\frac{r^2(h)dh}{\lambda+r(h)} +
\frac{(m+\lambda)\lambda}{\int_0^H r(h) dh} \int_0^H\frac{r(h)dh}{\lambda+r(h)}=0,\label{chareq3}
\end{align}

Let us first consider the case of a small  environment, i.e. small $H$.  We use the Taylor expansion of \eqref{chareq3}. We assume that the derivative of $r(h)$ at $h=0$ does not vanish: $r(h)\approx r_0+Rh$, where $R \ne 0$. The characteristic equation becomes
\begin{equation}
0=\frac{r_0 m+\lambda^2}{\lambda+r_0}+\frac{1}{2}r_0\gamma \lambda R \frac{m-\lambda}{(\lambda+r_0)^2}H-\frac{1}{12r_0} R^2\lambda \frac{r_0 m -3 \lambda m -3r_0 \lambda + \lambda^2}{(\lambda+r_0)^3}H^2+O(H^3) \label{expansion}.
\end{equation}
This expression can be re-written in the polynomial form ($\lambda \ne -r_0$) by dropping the term $O(H^3)$
\begin{equation}
0=a_0+a_1\lambda+a_2\lambda^2+ a_3\lambda^3 +a_4 \lambda^4,\label{polynom}
\end{equation}
where $a_0=r_0^3m$, $a_1=m(2 r_0^2-H^2 R^2/12+H r_0 R/2)$, $a_2=m r_0 +r_0^2 -(r_0-m)RH/2-R^2(3m+3r_0)H^2/(12r_0)$, $a_3=2r_0-RH/2-R^2H^2/(12r_0)$, $a_4=1$.

We use the Routh - Hurwitz stability criterion for the coefficients of the above polynomial: $a_n>0$, $a_2a_3>a_4a_1$, $a_1a_2a_3>a_4a_1^2+a_0a_3^2$.
One can see that for a sufficiently small $H$: $a_n>0$ ($n=1,2,3,4$), moreover
\begin{align}
a_1a_2a_3-a_4a_1^2-a_0a_3^2 &=m r_0^2(r_0+m)R^2H^2/2+O(H^3)>0, \label{coeff_1} \\
a_2a_3-a_4a_1 &= 2 r_0^3+O(H)>0.
\end{align}
Thus, for small $H$ the polynomial characteristic equation \eqref{polynom} has eigenvalues with only negative real parts and, consequently, the initial characteristic equation \eqref{chareq3} also exhibits stability for small $H$. Note that stability is sustained in the case when we consider a parabolic  approximation $r(h)\approx r_0+Rh+R_1 h^2$, where $R \ne 0, R_1 \ne 0$ (the need for considering such an approximation comes from the fact that in the second order expansion \eqref{expansion} the quadratic term $R_1$ will be present.  We should stress that here we are allowed to drop $O(H^3)$ and consider \eqref{polynom} since all the roots of \eqref{polynom} are perturbations of the solutions with $H=0$ such that the maximal order of perturbation with respect to $H$ does not exceed 2, i.e. it is less than 3 (we do not show the corresponding expressions $\lambda(H)$ of solutions of \eqref{polynom} for the sake of brevity).

Now let us suggest that for some critical $H>0$ the stability of \eqref{chareq3} changes and the real part of an eigenvalue becomes positive. This can occur at a Hopf bifurcation point and the condition of stability change is $\lambda=ib$, where $b$ is a positive real number. We substitute $\lambda=ib$ into the complex characteristic equation and obtain two equations for its real and imaginary parts. After some re-arrangement the equation for $b$ becomes (see electronic Appendix 2 for the exact derivation)
 \begin{align}
& -2\frac{\int_0^H r(h) dh}{\int_0^H r^2(h) dh}+\frac{\int_0^H\frac{r(h)dh}{b^2+r^2(h)}}{\int_0^H\frac{r^2(h)dh}{b^2+r^2(h)}}\left(1+ \frac{b^2 \int_0^H\frac{r(h)dh}{b^2+r^2(h)} }{\int_0^H r(h) dh}\right) + \frac{\int_0^H\frac{r(h)dh}{b^2+r^2(h)}}{\int_0^H r(h) dh} =0.\label{chareq4}
\end{align}
We substitute a particular parametrization $r(h)$ to verify whether or not \eqref{chareq4} has a real value solution. In the case \eqref{chareq4} does not have real solutions, the steady state of \eqref{prey1}-\eqref{predator1} will be always locally stable for $\beta=0$ because for small $H$ the eigenvalues always have non-zero negative real parts and for any $H>0$ they will not vanish (otherwise we would have a solution with $\lambda=ib$, in which case $b$ would solve \eqref{chareq4}).

 However, it is still an open question whether or not the equilibrium will remain stable for $\beta=0$ for any arbitrary positive function $r(h)$. In our investigation, we checked several concrete families of ecologically relevant and mathematically tractable functions including linear, parabolic, cubic, exponential and sinusoidal functions given, respectively, by $r(h)=a_0+a_1h$, $r(h)=a_0+a_2h^2$, $r(h)=a_0+a_3h^3$, $r(h)=a_0\exp(a_4 h)$ and $r(h)=a_0(1+\epsilon \sin(\omega h))$, where $a_i, \omega, \epsilon$ are parameters. For those functions we can obtain analytical expressions for the corresponding integral and numerically check the possibility to find a solution of \eqref{chareq4}, by varying the parameters. For all these functions we find that \eqref{chareq4} has no real solution, thus the equilibrium is always stable (we do not show the corresponding cumbersome expressions and graphs here for the sake of brevity).

%

\subsection{Model 2: local growth rate of prey is affected by self-shading }


\subsubsection{Existence of a non-trivial (coexistence) steady state}\

We look for the coexistence equilibrium $(p_*,Z_*)$ of model \eqref{prey2}-\eqref{predator2}.
The steady state equations read
\begin{align}
r_0 \exp\left(-\int_0^h \nu p(h,t)\ud h\right) & = \alpha\frac{Z_*}{P_*}\frac{p_*(h)}{1+\beta p_*(h)}, \label{genss21} \\
m  & = \frac{\alpha k}{P_* H}\int_0^H\frac{p_*^2(h)}{1+\beta p_*(h)}\,\ud h,  \label{genss22}\\
P_* & = \frac{1}{H}\int_0^H p_*(h)\,\ud h.
\end{align}
From the first equation of the system we obtain
\begin{align}
\ln(r_0) - \nu \int_0^h  p(h,t)\ud h & = \ln \left(\alpha\frac{Z_*}{P_*}\frac{p_*(h)}{1+\beta p_*(h)}\right). \label{genss23}
\end{align}
We differentiate the above expression to obtain
\begin{align}
-\nu & = \frac{1}{p_*^2(h)(1+\beta p_*(h))}p_*^\prime(h), \label{station_diff}
\end{align}
Integration of \eqref{station_diff} gives
\begin{align}
-\nu h+C & = -\frac{1}{p_*(h)} +\beta \ln \left( \frac {1+\beta p_*(h)}{p_*(h)} \right), \label{station_int}
\end{align}
where $C$ is a constant of integration. The constant $C$ can be found by setting $h=0$ and using  the expression for $p_*(0)$ from \eqref{genss21}
\begin{align}
p_*(0) & = \frac{P_* r_0}{\alpha Z_*-r_0 P_* \beta}, \label{p(0)}
\end{align}
After substituting $C$ and some simplification, we obtain the expression for the stationary state $p_*(h)$:
\begin{align}
-\nu h+\frac{r_0 P_*\beta-\alpha Z_*}{r_0 P_*} +\beta \ln \left(\frac{\alpha Z_*}{r_0 P_*}\right)& = -\frac{1}{p_*(h)} +\beta \ln \left( \frac {1+\beta p_*(h)}{p_*(h)} \right), \label{station_int_final}
\end{align}
The above expression is implicit since it is impossible to solve it analytically for $p_*(h)$ except in the particular case $\beta=0$; however, from \eqref{station_diff} one can conclude that $p_*(h)$ is a decreasing function of $h$. The maximal density through the column is attained at $h=0$ and given by \eqref{p(0)}; the minimal density is achieved at $h=H$. From \eqref{genss21} we obtain the expression for $p_*(H)$
\begin{align}
p_*(H) & = \frac{P_* r_0 \exp(-H\nu P_*)}{\alpha Z_*-r_0 \exp(-H\nu P_*) P_* \beta}, \label{p(H)}
\end{align}
We use \eqref{p(H)} and  \eqref{station_int_final} to get the following relation between $P_*$ and $Z_*$
\begin{align}
0 & = -\exp(P_* H \nu) \alpha Z_*+\beta P_*^2 H \nu+ H \nu r_0 P_* +\alpha Z_*. \label{Z_and_P}
\end{align}

Next we consider the stationary equation \eqref{genss22} for the predator and replace the integration variable $h$  with $p(h)$ using \eqref{station_diff}.
\begin{align}
m & = \frac{\alpha k}{P_* H}\int_0^H\frac{p_*^2(h)}{1+\beta p_*(h)}\,\ud h = -\frac{\alpha k}{P_* H \nu}\int_{p_*(0)}^{p_*(H)}\frac{1}{(1+\beta p_*(h))^2}\,\ud p=\frac{\alpha k}{P_* H \nu \beta} \left[
  \frac {1}{1+\beta p} \right]_{p_*(0)}^{p_*(H)}. \label{genss23}
\end{align}
We are allowed to make the above change of variables since $p(h)$ is a monotonically decreasing function. We substitute the values of $p_*(0)$ and $p_*(H)$ from \eqref{p(0)} and  \eqref{p(H)}, respectively, to obtain after simplification another identity linking $P_*$ and $Z_*$
\begin{align}
Z_* & = \frac{(1-\exp(-P_* H \nu) )r_0 k}{m H \nu}. \label{eq_Z}
\end{align}
Finally, we substitute the expression for $Z_*$ from \eqref{eq_Z} into \eqref{Z_and_P} to obtain the equation for $P_*$ which reads
\begin{align}
 f(P_*) & = \beta P_*^2 H \nu + P_* H \nu r_0 +\frac{\alpha k r_0}{m H \nu} (\exp(- P_* H \nu)-1) (\exp( P_* H \nu)-1)=0, \label{eq_P*}
\end{align}
This is a transcendental equation; however, we can make exhaustive  conclusions about the existence of a  positive solution of \eqref{eq_P*}. At low $P_*$, the function $f(P_*)$ is linear and always has a positive slope. At large $P_*$, $f(P_*)$ becomes negative (the exponential term becomes dominant), thus there should be at least one point where this function vanishes. Additionally, equation \eqref{eq_P*} should be considered together with the condition $p_*(h)>0$, which is equivalent to $p_*(0)>0$ since $p_*(h)$ is a decreasing function. From \eqref{p(0)} and \eqref{eq_Z} it follows that the condition $p_*(0)>0$ is equivalent to
\begin{align}
 f_1(P_*) & = P_* \beta m H \nu +\alpha k \exp(-P_* H \nu)-\alpha k>0, \label{eq_P>0}
\end{align}
It is easy to see that \eqref{eq_P>0} will be satisfied for some $P_*>0$ if and only if $\alpha k - \beta m >0 $. Further, by computing the second derivative of $f(P_*)$ one can show that for $\alpha k - \beta m >0 $ it is always negative, thus there can be only one positive solution of \eqref{eq_P*}. Hence, combining the above properties of $f(P_*)$ and $f_1(P_*)$, we find that the existence of the nontrivial steady state of the system \eqref{prey2}-\eqref{predator2} is given by the following
\begin{proposition}\label{ssexistence}
Model \eqref{prey2}-\eqref{predator2} admits a unique positive steady state $(p_*,Z_*)$ in the case $\alpha k - \beta m >0 $ and the unique positive solution of \eqref{eq_P*}, which would always exist under the mentioned condition, satisfies the inequality \eqref{eq_P>0}.
\end{proposition}

It is easy to see that for $\beta \ll 1$ the conditions of the proposition are satisfied, thus there always exists a non-trivial steady state. Let us consider a gradual increase in $\beta$ while still keeping $\alpha k - \beta m >0 $.  The root of $f(P_*)=0$ will increase, whereas the root of $f_1(P_*)=0$ will decrease. At a certain $\beta_*$ both roots coincide, which is an unavoidable event since the root of $f_1(P_*)=0$ will approach zero  when $\alpha k = \beta m  $. Thus for all $\beta<\beta_*$, the system has a non-trivial coexistence equilibrium whereas for  $\beta>\beta_*$, a positive equilibrium is impossible. The exact value of $\beta_*$ can be found from the condition $f(P_*)=f_1(P_*)=0$ which results in a cumbersome analytical expression (not shown here).

\subsubsection{Stability of the non-trivial (coexistence) steady state of Model 2}\

The linearisation around the positive steady state $(p_*,Z_*)$ of \eqref{prey2}-\eqref{predator2} results in a rather cumbersome characteristic equation.
In this paper we focus on an important particular case, where there is no saturation in the functional response of predator, i.e. $\beta=0$. The linearized system becomes
\begin{align}
\frac{\partial \Delta p}{\partial t}(h,t) =&r_0 \exp\left(-\int_0^h \nu p_*(h)\ud h\right) \left[ \Delta p(h,t)-\nu p_*(h) \int_0^h \Delta p(x,t)\ud x \right] \nonumber \\
 & -\alpha\left(\frac{p^2_*(h)}{P_*}\Delta Z(t)-\frac{p^2_*(h)Z_*}{P_*^2}\Delta P(t)+\frac{2Z_*p_*(h)}{P_*}\Delta p(h,t)\right), \label{linear_ppM2} \\
\frac{\ud \Delta Z}{\ud t}(t) =& -\frac{\alpha k Z_*}{HP_*^2}\Delta P(t)\int_0^H p^2_*(h)\,\ud h + \frac{\alpha kZ_*}{HP_*}\int_0^H 2 p_*(h) \Delta p(h,t)\,\ud h. \label{linear_zzM2}
\end{align}
 We look for the solution in the form of $\Delta p (h,t)=w(h)\exp(\lambda t)$ and $\Delta Z (t)=U\exp(\lambda t)$. We substitute the expressions for the stationary solutions \eqref{station_int}  ($\beta=0$) and \eqref{eq_Z} to arrive at the following eigenvalue problem ($W$ is the spatial average of $w$)
\begin{align}
w(h)\lambda =&\frac{C r_0}{h \nu +C}\left[ w(h)-\nu \frac{1}{h \nu +C}\int_0^h w(x)\ud x \right] \nonumber \\
 & -\frac{C r_0}{(h \nu +C)^2}\left(\frac{U}{Z_*}-\frac{W}{P_*}+2(h \nu +C) w(h)\right), \label{eigen_p_M2} \\
 U \lambda=& \frac{ k C r_0}{H}\left( -\frac{1}{P_*}W\int_0^H \frac{1}{(h \nu +C)^2}\,\ud h + 2 \int_0^H  \frac{w(h)}{h \nu +C} \,\ud h \right). \label{eigen_z_M2}
\end{align}
We re-arrange the first equation of the above system
\begin{align}
w(h)\lambda (h \nu +C)^2 =& -C r_0 (h \nu +C)w(h) -\nu C r_0 \int_0^h w(x)\ud x -C r_0 \left(\frac{U}{Z_*}-\frac{W}{P_*}\right), \label{eigen_pp_M2}
\end{align}
We differentiate both sides of \eqref{eigen_pp_M2} and obtain
\begin{align}
\left( \frac{d w}{dh} (h \nu +C) + 2w(h)\nu \right)\lambda (h \nu +C)=& C r_0 \left(-\nu w(h) - \frac{d w}{dh} (h \nu +C)-\nu w(x)\right) \label{diff_pp_M2}
\end{align}
After simplifying the above equation we get
\begin{align}
\left( (h \nu +C)\lambda +\nu C \right)(h \nu +C)\frac{d w}{dh}=& -2 \nu w(h)\left( (h \nu +C)\lambda +\nu C \right) \label{diff_eq_M2}
\end{align}
We suggest that $(h \nu +C)\lambda +\nu C  \ne 0$ and derive the following equation for the eigenfunction $w(h)$
\begin{align}
(h \nu +C)\frac{d w}{dh}=& -2 \nu w(h) \label{diff_eq1_M2}
\end{align}
Integration of this linear equation gives the following eigenfunction
\begin{align}
 w(h)=\frac{1}{(h \nu +C)^2}. \label{w(h_M2}
\end{align}
Next, we substitute this function into system \eqref{eigen_p_M2}-\eqref{eigen_z_M2}
\begin{align}
\lambda =& -r_0 -C r_0\left(\frac{U}{Z_*}-\frac{W}{P_*}\right), \label{eigen_p1_M2} \\
 U \lambda=& \frac{ k C r_0}{H}\left( -\frac{1}{P_*}W\int_0^H \frac{1}{(h \nu +C)^2}\,\ud h + 2 \int_0^H  \frac{1}{(h \nu +C)^3} \,\ud h \right), \label{eigen_z1_M2}
\end{align}
where the value of $W$ is obtained from
\begin{align}
W=& \frac{1}{H}\int_0^H {w(h)} dh = \frac{1}{H} \int_0^H \frac{dh}{(h \nu +C)^2}=\frac{1}{C(H\nu+C)}.  \label{W_M2}
\end{align}
We simplify equation \eqref{eigen_z1_M2}
\begin{align}
 U \lambda=& \frac{ k r_0}{(h \nu +C)^2 C}\left( -\frac{1}{P_*} + 2C+H\nu \right), \label{eigen_z11_M2}
\end{align}
finally, we substitute the expressions for the stationary value $P_*$ into \eqref{eigen_p1_M2} and \eqref{eigen_z11_M2} to obtain the characteristic equation
\begin{align}
0=\lambda^2+ & (1 -Q_1)r_0 \lambda + Q_2, \label{eigen_complete_1_M2}
\end{align}
where
\begin{align}
Q_1 = \frac{H\nu}{(C+H \nu)\ln((C+H\nu)/C)}< 1,
\end{align}
\begin{align}
Q_2 =\frac{ k r_0}{(h \nu +C)^2 C Z_*}\left( -\frac{H\nu }{\ln((C+H\nu)/C)}+ 2C+H\nu \right)>0.
\end{align}
It is easy to see that characteristic equation \eqref{eigen_complete_1_M2} always has negative real parts, thus, the linearized system \eqref{linear_ppM2}-\eqref{linear_zzM2} is always stable.

We should also say that unlike the model without self-saturation the number of eigenvalues is finite and there is a single  eigenfunction. This can be explained by the fact that separation of variables provides us with only the solution for large values of time and does not describe transient regimes at small and intermediate times.

The above result can be summarized as the following
\begin{theorem}
The steady state $(p_*,Z_*)$ of model \eqref{prey2}-\eqref{predator2} is always locally asymptotically stable for $\beta=0$ (Holling type I functional response).
\end{theorem}

\begin{remark}
 One can prove that in a similar way the stability of equation  \eqref{linear_ppM2} governing  the dynamics of prey for a constant (equilibrium) density of predator, i.e. for $Z(t)=const$ or $U(t)=0$. The eigenvalue of the system is given by
\begin{align}
\lambda= & -(1 -Q_1)r_0 <0.
\end{align}
  This signifies that a spatial regrouping of the predator population (without variation in the total population size) alone can control the prey population in the case of self-shading.
\end{remark}

\section{Discussion and summary of results}
In this paper, we revisit predator-prey interaction in ecosystems with a high nutrient load and explore the role that spatial heterogeneity of the environment and high predator mobility have in stability of such  ecosystems. Our analytical investigations of the model predict that interplay between these two factors results in stabilization of an otherwise globally unstable predator-prey system. An important assumption for the realization of the given  stabilization scenario is that the predator preferentially feeds at locations with high prey density (see relationship \eqref{z_distribution}). Since spatial gradients in a species' growth rate across their habitats is a typical ecological situation, and many  predators/consumers are highly mobile and feed mainly in patches of high food density \cite{Milinski1979, Godin1984, Lodge1988, Kacelnik1992, Begon2005, Nachman2006} (see also figure 1 of the current paper), we claim here that the reported mechanism should be rather typical for marine, freshwater and terrestrial ecosystems under eutrophication. Our findings also provide a new and robust mechanism to resolve the famous 'paradox of enrichment', which is a long standing enigma in theoretical ecology \cite{Rosenzweig_1971, Gilpin1972, Abrams_walters, Genkai-Kato_1999, Petrovskii2004, Mougi_Nishimura2007, Roy2007}.

Our models are formulated as integro-differential equations, which allows for the incorporation of different time scales: fast spatial movement of predator and slow population dynamics of both species. This makes the current framework more effective than classical reaction-diffusion models for modelling trophic interaction in small-sized habitats, since individuals cannot be assigned to a particular spatial location in this case while reaction-diffusion equations contain only local terms.

Here we explore a generic predator-prey model with an infinitely large carrying capacity of prey (i.e. no resource limitation) and consider two main scenarios: (i) the prey growth rate gradient is fixed (Model 1) and the local growth rate of prey is affected by self-shading (Model 2). The main mathematical outcomes of our study are the following:

(i) For each model (Model 1 and Model 2) we find the condition for the existence of the positive stationary state through which prey and predator can coexist for an arbitrary gradient of the growth rate $r(h)\ge 0$. Interestingly, including self-shading usually restricts the conditions of coexistence (cf. Propositions 3.1 and 3.5).

(ii) For Model 1 we have derived the general characteristic equation \eqref{chareq} governing the stability of the coexistence equilibrium. Using this equation, one can construct the Hopf bifurcation curve that separates stable and unstable dynamics without numerically integrating the underlying partial integro-differential equations (see an example below).

(iii) Analysis of the characteristic equation \eqref{chareq} shows that for a homogeneous environment, in which $r(h) \equiv  const$, the system is always unstable provided there is saturation in the functional response (Remark 3.4). Thus, spatial heterogeneity of the environment is a necessary condition for stability of the coexistence equilibrium.

(iv) For both Model 1 and Model 2, in the case of a linear (Holling type I) functional response of the predator, we analytically proved that the coexistence equilibrium is always stable. Since the system is structurally stable, the stability of the equilibrium is also guaranteed for sufficiently small $\beta>0$, i.e. for a Holling type II functional response with low saturation.

(v) Comparison of Model 1 and Model 2 shows that stabilization in the system with dynamical self-shading (Model 2) is potentially a stronger mechanism than the scenario, where the spatial gradient in the prey growth rate is fixed (Model 1). For instance, without saturation in the functional response ($\beta=0$) the stabilization does not require changing the predator biomass: variation in the spatial distribution of a fixed number of predators can suppress the prey outbreak and re-establish the equilibrium (cf. Remarks 3.3 and 3.7).

In the case of a fixed prey growth rate gradient, the general characteristic equation \eqref{chareq} allows us to easily compute the Hopf bifurcation curve to direct numerical simulation of the underlaying integro-differential equations \eqref{prey1}-\eqref{predator1}. As an illustrative example, here we have constructed the Hopf bifurcation curve for the system with a linear variation of the growth rate $r(h)=a_0+a_1h$ (see electronic Appendix 3 for detail). Fig.2 shows a set of bifurcation curves in the diagram $a_1$ and $\beta$ constructed for a gradually increasing size $H$ of the habitat. In each case, the stability region is located below the curve. One can see that increased the saturation in food consumption, signified by a large $\beta$, usually acts as a destabilizing factor, but the influence of the gradient of the growth rate $a_1$ is more complicated: small and large $a_1$ result in destabilization of the system whereas stabilization occurs for some intermediate $a_1$. An increase in the size of the habitat can also either destabilize or stabilize the system. This result is somewhat counterintuitive since in earlier works it was shown that for an exponential parametrization of $r(h)$, an increase in the spatial gradient of the growth function or the size of the habitat will always facilitate stabilization of the system \cite{Morozov_2011, Lewis2013}.

Our analytical investigation confirms the numerical results reported earlier in \cite{Morozov_2011, Lewis2013} about the possibility of the stabilization of systems with 'unlimited carrying capacity', and extend the previous findings to more general parametrizations of the spatial gradient of $r(h)$. In particular, we proved that for a small habitat size $H$, any arbitrary non-homogeneous dependence $r(h)$ will stabilize the predator-prey interactions (see Section 3.1.3). On the other hand, our work shows that using some other parametrizations, for example a linear function, can provide some qualitatively different prediction compared to the exponential dependence considered in \cite{Morozov_2011, Lewis2013}. This underlines the danger of sticking to particular parametrizations when modelling predator-prey systems.

Throughout the paper we have neglected the diffusion of the prey population by setting $D=0$ in equation \eqref{prey}. Including diffusion $D>0$ will make the model more realistic, but the equations become analytically untractable. Although one can still perform linearization techniques, it becomes impossible to obtain the resultant characteristic equation due to the presence of the second order derivative with respect to $x$,  thus the only way of exploring the system is by direct numerical simulation. For small values of $D$ the main conclusions of our paper obtained for $D=0$ will hold due to the  structural stability of the  model equations. Interestingly, for some ecosystems (e.g. planktonic systems, \cite{Morozov_2011, Lewis2013}) the diffusion coefficient $D>0$ is relatively small (compared to the interaction terms) so the results obtained should be close to those with $D=0$ . However, for a very large diffusion coefficient corresponding to a highly mobile prey, the system becomes homogeneous in space and is reduced to the classical Rosenzweig-MacArthur model, which is always globally unstable for any $\beta>0$. Thus, the system will be destabilized for a certain critical $D$, which can be found numerically.

Finally, we should mentioned the open questions that our study raises. For instance, it would be an important extension to determine the stability or otherwise of the coexistence equilibrium of Model 1 in the case $r(h)$ is an arbitrary function for a Holling type I functional response (see equation \eqref{chareq4}). Based on preliminary results obtained for a small habitat size $H$, we hypothesize that for $\beta=0$ the Model 1 should be stable for an arbitrary $r(h)$, but it remains to prove this conjecture rigorously.  Another challenging issue is exploring the system stability for a more complicated relationship between $z(h,t)$ and $p(h,t)$, by taking into account possible interference between predators \cite{Murdoch1992}. For instance, this can be represented by setting $z(h)=\frac{p(h)^\mu}{P^\mu}Z, \mu>0$ \cite{Murdoch1992}. Another interesting extension would be to investigate analytically the model combining self-shading of the population with a fixed gradient of local growth rate and combining Model 1 and Model 2. From the biological point of view it might even be natural to consider situations when the growth function $R$ (or $r$) can take negative values, i.e. the population of prey actually declines locally.
In this more general case, model \eqref{prey1}-\eqref{predator1} will still be well-posed, and solutions starting in the positive cone would remain non-negative for all times. At the same time equation \eqref{steady1} shows that there will be no strictly positive steady states of the model and this might influence the previous stability results. Models with locally negative growth rate of prey could therefore possess quite  rich dynamics, and such models should definitely be a matter of further investigation.

\begin{figure}[H]
\centering
\includegraphics[width=13cm]{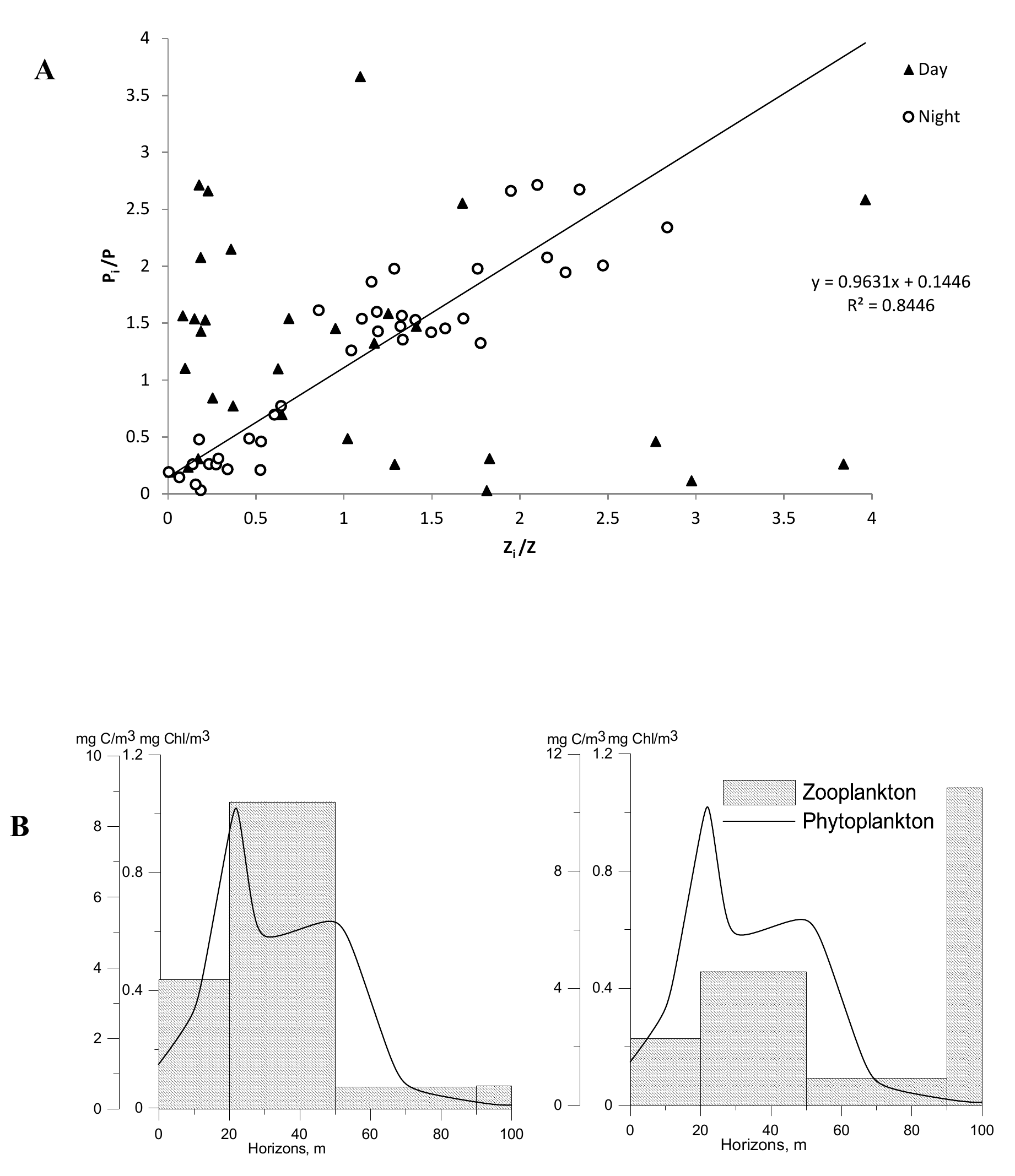}  
\caption{(A) Relative distribution of phytoplankton and zooplankton biomass across the water column collected at 12 stations in the Black Sea. Details on plankton sampling are provided in electronic Appendix 1. For each station, the whole euphotic zone is split into a number of layers. Figure shows the relative phytoplankton biomass  plotted against the relative zooplankton biomass  in the same layer. Here $P_i$ and $Z_i$ are the average densities of species in layer $i$; $P$ and $Z$ are the overall average densities. \newline
(B) Typical example of vertical distribution of zooplankton ($mg C/m^3$) and phytoplankton ($mg Chl/m^3$) during night (bottom left panel) and daytime (bttom right panel) collected at station \#5. The profile of phytoplankton distribution is averaged over 24 h. High variation in zooplankton distribution through the day is due to the diel vertical migration (for more explanation see text).}
\label{Figure1}
\end{figure}

\begin{figure}[H]
\centering
\includegraphics[width=13cm]{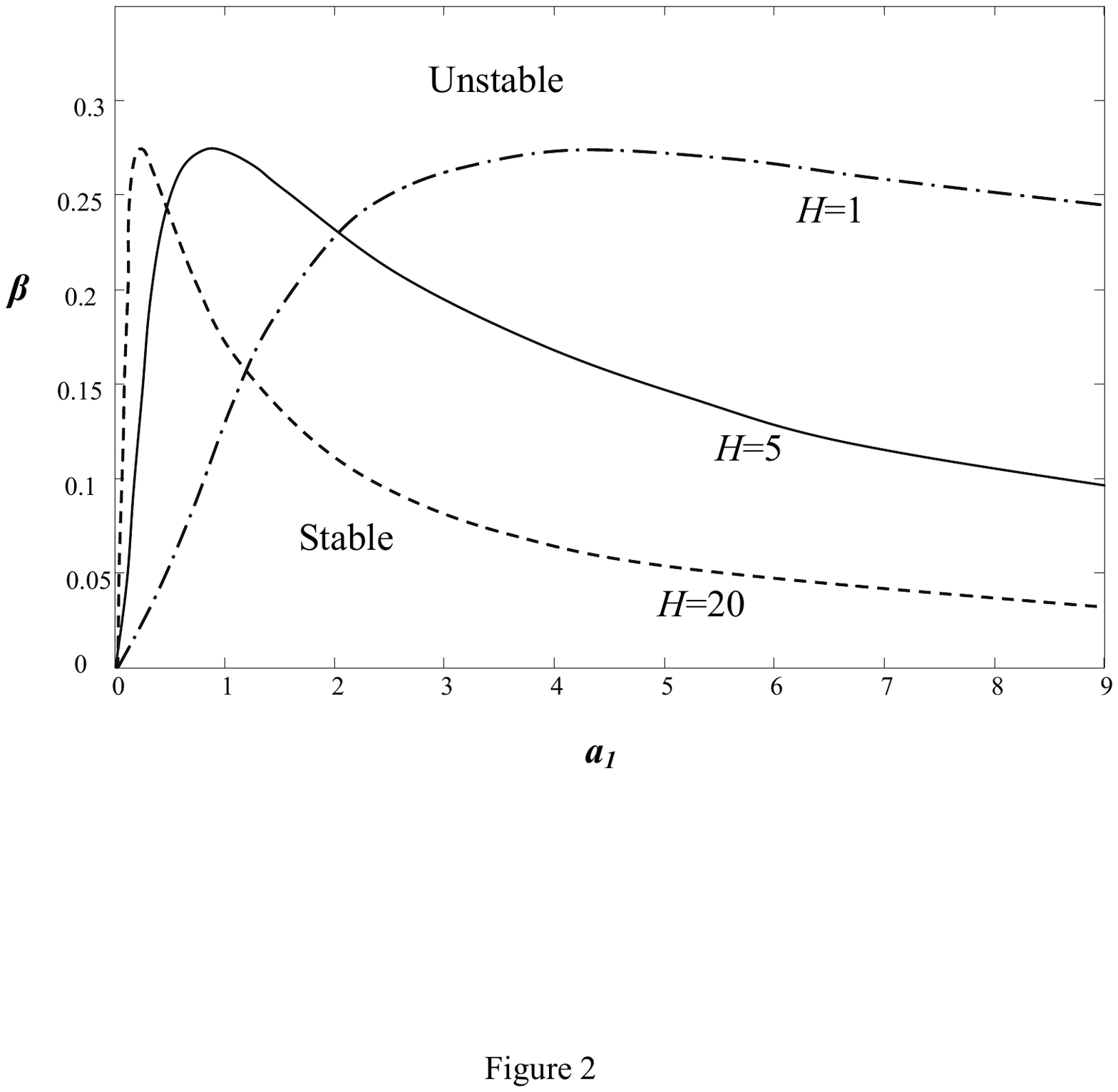}  
\caption{Hopf bifurcation curves constructed for model (2.6)-(2.7) for the case the growth rate is a linear function of space $r(h)=a_0+a_1 h$. The curves are plotted using the derived characteristic equation (3.34), for details see Appendix 3. Various curves are obtained for different size $H$ of the habitat and each of them separates the stability region (below the curve) from the instability region (above the curve). The other model parameters are $m=0.1; a_0=1; k=1; α=2$.}
\label{Figure2}
\end{figure}

\section*{Acknowledgments}

We thank the anonymous Reviewer for his/her helpful comments and suggestions.

\end{document}